\documentclass[12pt]{amsart}

\newcommand{\CC}{\mathbb C}
\newcommand{\Z}{\mathbb Z}
\newcommand{\Ker}{\mathop{\rm Ker}\nolimits}

\begin{document}
\title[Integral of exponent of a polynomial]
{Integral of exponent of a polynomial is a generalized hypergeometric function of the coefficients of the
polynomial}
\begin{abstract}
We show that the integral $\int e^{S(x_1,\ldots,x_n)}dx_1\ldots dx_n$ for an arbitrary polynomial $S$,
satisfies a generalized hypergeometric system of differential equations in the sense of I.~M.~Gelfand et al.
\end{abstract}
\thanks{Partially supported by the grant RFBR 10-01-00536}

\author{A. V. Stoyanovsky}

\address{Russian State University of Humanities}
\email{alexander.stoyanovsky@gmail.com}

\maketitle

Integrating over a linear space with exponential weight is an important procedure in statistical and
quantum physics. However, till recent time, the integral of exponent of a polynomial of several variables has been
known only for quadratic polynomials (Gaussian integral).

The pioneering works on integrating exponents of non-quadratic polynomials are due to V.~V.~Dolotin, A.~Yu.~Morozov,
and Sh.~R.~Shakirov [1--5]. In these works one considers homogeneous polynomials (forms). In the paper [5],
in order to compute integral of exponent of a form, one uses differential equations satisfied by the integral
as a function of the coefficients of the form, and one also uses invariant theory. It is shown that
in several particular cases, the integral is a generalized hypergeometric function in the sense of
I.~M.~Gelfand {\it et al.} [6] of algebraic invariants of a form. The authors of [1--5]
consider the integral as an invariant of the special linear group, and use the term ``integral discriminant''
for it.

The purpose of the present note is to make the following simple observation: the system of differential equations
satisfied by the integral of exponent of an arbitrary (not necessarily homogeneous) polynomial, as a function
of the coefficients of the polynomial, coincides with the generalized hypergeometric system (GHS) [6],
so that the integral is a generalized hypergeometric function of the coefficients of the polynomial.

Indeed, consider the integral
\begin{equation}
Z=\int e^{S(x)}dx,
\end{equation}
where $x=(x_1,\ldots,x_n)$, $dx=dx_1\ldots dx_n$,
\begin{equation}
S(x)=\sum_{k\in K}c_kx^k,
\end{equation}
$k=(k_1,\ldots,k_n)$, $x^k=x_1^{k_1}\ldots x_n^{k_n}$, $S(x)$ is a polynomial with complex coefficients, and
integration goes over an $n$-dimensional real contour in $\CC^n$, on which the function $e^{S(x)}$
rapidly decreases at infinity. Assume that the degrees of the monomials $k\in K$ span the space $\CC^n$.
One has a surjective linear map
$\pi:\CC^K\to\CC^n$, $\sum_{k\in K}n_ke_k\mapsto\sum_{k\in K}n_k\cdot k$, where $e_k$ is the basis vector in $\CC^K$
corresponding to $k$. Consider the lattice $B=\Z^K\cap\Ker\pi\subset\CC^K$, the subspace
$L=\Ker\pi$ spanned by $B$, and the vector $\alpha=(-1,-1,\ldots,-1)\in\CC^n\simeq\CC^K/L$. Let $A\subset(\CC^K)'$
be the annihilator of $L$, $A\simeq(\CC^n)'$.

\medskip

{\bf Theorem.} {\it The integral $(1)$
satisfies the generalized hypergeometric system of equations on the space $\CC^K$
corresponding to the lattice $B$ and to the parameter $\alpha$.
}
\medskip

{\it Proof.} Recall that GHS reads ([6], formulas (0.5),(0.6))
\begin{equation}
\prod_{k:n_k>0}\left(\frac\partial{\partial c_k}\right)^{n_k}Z=
\prod_{k:n_k<0}\left(\frac\partial{\partial c_k}\right)^{-n_k}Z\text{ for all }(n_k)\in B,
\end{equation}
\begin{equation}
\sum_{k\in K}a_kc_k\frac{\partial Z}{\partial c_k}=\langle a,\alpha\rangle Z\text{ for all }a\in A.
\end{equation}
Equality (3) is obvious. To check (4), it suffices to put $a_k=k_i$ for a fixed $i$, $1\le i\le n$.
Then the left hand side of (4) equals
$$
\int\sum_k k_ic_kx^ke^{S(x)}dx=\int x_i\frac{\partial}{\partial x_i}e^{S(x)}dx=-\int e^{S(x)}dx
$$
(integration by parts), Q. E. D.

The author is grateful to V. V. Dolotin for numerous discussions, and to Sh.~R.~Shakirov
for his talk about the paper [5] at the author's seminar at the Independent Moscow University.

{\it Remark.} It turned out that in the paper [7], I.~M.~Gelfand and M.~I.~Graev pointed out
that integral of exponent of a sum of monomials with arbitrary complex powers of
variables formally satisfies generalized hypergeometric system of equations, so our result is not new
in this respect.
An essential difference with our setup is that the authors of [7] consider integration contours in $(\CC\setminus0)^n$
which do not go to infinity. In the case of exponent of a usual polynomial, integral over such contour vanishes.

\end{document}